\documentstyle[prl,aps,epsf]{revtex}
\begin{document}
\draft

\twocolumn

\narrowtext
\noindent{\bf Comment on ``Effects of Point Defects on the Phase Diagram of Vortex
States in High-$T_c$ Superconductors in the B $||$ c Axis''}

$\,$

Recently Nonomura and Hu (N-H) presented \cite{R1} simulations of a bond disordered
three dimensional uniformly frustrated XY model, as a model for
a point disordered type-II superconductor, using very strongly anisotropic 
couplings $J_c/J_{ab}=(\lambda_{ab}/\lambda_c)^2=1/400$.  
As part of their phase diagram, they report a {\it vortex slush} (VS) phase
that lies at disorder strengths $\epsilon$ above the elastically distorted vortex line lattice 
(the {\it Bragg glass} (BG)) and is separated from it by a first order phase transition 
$\epsilon_{c1}(T)$.  Increasing either disorder strength or temperature, the VS
has another first order transition $\epsilon_{c2}(T)$ to the vortex liquid (VL).
The VS 
has neither superconducting phase coherence (longitudinal helicity modulus vanishes, $\Upsilon_c=0$)
nor translation order (liquid-like structure function $S({\bf k})$), and is
distinguished from the VL by a sharp drop in the density of dislocations
in the $ab$ plane.
Here we clarify the nature of the VS found by N-H, 
and argue that it is the result of an unphysical finite size effect.  

\begin{figure}
\epsfxsize=7.5truecm
\epsfbox{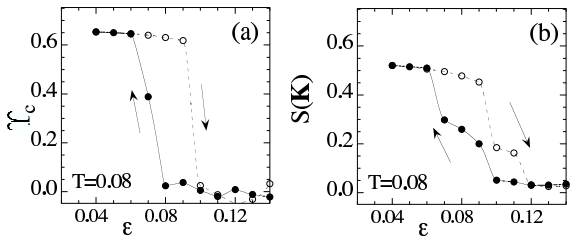}
\caption{a) Helicity modulus $\Upsilon_c$ and b) 
structure function peak height $S({\bf K})$ vs. disorder strength $\epsilon$
at $T=0.08$. Solid (open) symbols are for decreasing (increasing) $\epsilon$.}
\label{f1}
\end{figure}

We have carried out simulations of N-H's model, using all the same parameters.  
We first cool to $T=0.08$ at high $\epsilon=0.14$, staying within the VL.
 Then we slowly decrease $\epsilon$ to enter first the VS and then finally the BG.
 Then we slowly increase $\epsilon$ back to $0.14$.
In Fig.~\ref{f1}a we plot our results for the helicity modulus, $\Upsilon_c$.  
In Fig.~\ref{f1}b we plot $S({\bf K})$, where ${\bf K}$ 
is a reciprocal lattice vector of the vortex line lattice we find in the BG.
Decreasing $\epsilon$, we find that $\Upsilon_c$ rises from zero, indicating
the onset of superconducting phase coherence, and $S({\bf K})$ saturates to its low
$\epsilon$ value, indicating formation of an ordered vortex line lattice, both at 
$\epsilon_{c1}\sim 0.06$.  Increasing $\epsilon$, however, we find that both 
phase coherence and the vortex line
lattice persist to the higher value $\epsilon_{c2}\sim 0.10$.
Our values of $\epsilon_{c1}$ and $\epsilon_{c2}$ are reasonably close
(given likely sample to sample fluctuations) to the
values reported by N-H for the boundaries of the VS at $T=0.08$.
We therefore identify $\epsilon_{c1}<\epsilon<\epsilon_{c2}$ as the VS
and conclude that it is a region
with considerable hysteresis and metastability.

\begin{figure}
\epsfxsize=7.5truecm
\epsfbox{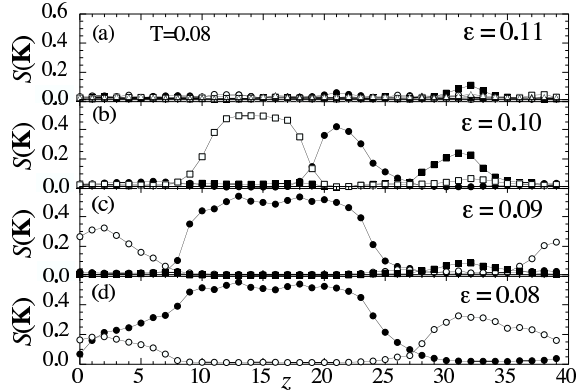}
\caption{Vortex structure function $S({\bf K})$ computed layer
by layer at heights $z$, for several different $\epsilon$, upon
decreasing $\epsilon$.  The different
symbols correspond to different values of ${\bf K}$, representing 
different vortex lattice orientations.}
\label{f2}
\end{figure}

To get a clearer picture of the VS, we plot in Fig.~\ref{f2}
$S({\bf K})$, computed layer by layer for the $ab$ planes
at different heights $z$, for several different values of $\epsilon$
as obtained when decreasing $\epsilon$. 
The different symbols represent different values of ${\bf K}$,
corresponding to vortex lattices of different orientation. We see
that in the VS, vortices in most planes have ordered into a 2D vortex
lattice, however the orientation of this lattice varies in different layers.
Decreasing $\epsilon$, the thickness of aligned layers increases, until
below $\epsilon_{c1}$ in the BG, a single orientation fills the system.

For a particular random realization and system size $L$, it is possible
that the above VS is indeed more stable than an ordered vortex line lattice for
a region of $\epsilon$.
However this cannot remain true as $L$ increases.
For an ordered vortex lattice of size $L$ and thickness $L_z\sim L$,
the most energy that can be gained by adjusting to a particular
set of random point pins scales as $\epsilon J_{ab}L^{3/2}$.  The energy loss
by having a mismatch of vortex lattices between two adjacent planes will scale as
$J_zL^2$.  For small $J_z$, fixed $L$, such an adjustment may be favored.  But as $L$
increases, the energy cost always wins out and the vortex lattices will align for
all layers.  The vortex slush region will disappear, being replaced by a 
vortex line lattice, with a single melting transition to the liquid.

This work was supported by the
Engineering Research Program of the Office of Basic Energy Sciences
at the Department of Energy grant DE-FG02-89ER14017 and the
Swedish Research Council, contract No. 621-2002-3975.  
Travel between Rochester and Ume{\aa} was supported by 
grants NSF INT-9901379 and STINT 99/976(00).

\noindent \,

\noindent P. Olsson$^a$ and S. Teitel$^b$

$^a$Department of Theoretical Physics

Ume{\aa} University, 901 87 Ume{\aa}, Sweden

$^b$Department of Physics and Astronomy

University of Rochester, Rochester, NY 14627

\noindent \,


\end{document}